\documentclass[openacc]{rstransa}
\usepackage[utf8]{inputenc}
\usepackage{amsmath}
\usepackage{amssymb}
\DeclareMathOperator{\Tr}{Tr}
\usepackage{epstopdf}
\usepackage{hyperref}
\usepackage{cite}

\begin{document}

\title{Director alignment at the nematic-isotropic interface: elastic anisotropy and active anchoring}

\author{
Rodrigo C. V. Coelho$^{1,2}$, Nuno A. M. Araújo$^{1,2}$ and Margarida M. Telo da Gama$^{1,2}$}

\address{$^{1}$Centro de Física Teórica e Computacional, Faculdade de Ciências, Universidade de Lisboa, 1749-016 Lisboa, Portugal.\\
$^{2}$Departamento de Física, Faculdade de Ciências,
Universidade de Lisboa, 1749-016 Lisboa, Portugal.}

\subject{liquid crystals, computational physics, active matter}


\corres{Rodrigo C. V. Coelho\\
\email{rcvcoelho@fc.ul.pt}}

\begin{abstract}
Activity in nematics drives interfacial flows that lead to preferential alignment that is tangential or planar for extensile systems (pushers) and perpendicular or homeotropic for contractile ones (pullers). This alignment is known as active anchoring and has been reported for a number of systems and described using active nematic hydrodynamic theories. The latter are based on the one-elastic constant approximation, i.e., they assume elastic isotropy of the underlying passive nematic. Real nematics, however, have different elastic constants, which lead to interfacial anchoring. In this paper, we consider elastic anisotropy in multiphase and multicomponent hydrodynamic models of active nematics and investigate the competition between the interfacial alignment driven by the elastic anisotropy of the passive nematic and the active anchoring. We start by considering systems with translational invariance to analyze the alignment at flat interfaces and, then, consider two-dimensional systems and active nematic droplets. We investigate the competition of the two types of anchoring over a wide range of the other parameters that characterize the system. The results of the simulations reveal that the active anchoring dominates except at very low activities, when the interfaces are static. In addition, we found that the elastic anisotropy does not affect the dynamics but changes the active length that becomes anisotropic.
\end{abstract}


\begin{fmtext}

\end{fmtext}


\maketitle

\section{Introduction}

Nematics are the simplest of liquid crystalline phases exhibiting long-range orientational order and positional disorder~\cite{p1995physics}. 
The nematic phase is uniaxial with the particles aligned along a particular direction, the nematic director. Away from surfaces 
or boundaries this direction is arbitrary as the free energy depends only on the degree of order, rather than on its direction.
Boundaries, however, break the translational invariance of the bulk and this in turn breaks the rotational invariance of the 
nematic as both the translational and orientational degrees of freedom are coupled. In particular, this is observed at free 
surfaces and is known as interfacial anchoring. 
       
Interfacial anchoring has been a topic of interest in passive nematics since the seminal work by de Gennes~\cite{doi:10.1080/15421407108082773} 
revealed that  it is due to the elastic anisotropy of the nematic, i.e. the difference between the splay, bend and twist elastic constants. Specifically, de Gennes has shown that depending on the values of the elastic constants the free interface will 
exhibit planar (parallel) or homeotropic (perpendicular) anchoring. In addition, the elastic anisotropy implies that there are
two correlation lengths that set the scales for the relaxation of perturbations in directions parallel or perpendicular to the 
bulk nematic orientation~\cite{doi:10.1080/15421407108082773, p1995physics}. 

At the level of the simplest de Gennes theory the elastic anisotropy is described by two elastic constants in three dimensions
(the splay and bend constants are identical) and by a single elastic constant (elastic isotropy) in two dimensions.
Higher order generalizations of de Gennes theory, as well as simulations and theories based on density functional theory of 
microscopic models of nematics do account for the observed full elastic anisotropy of real nematics, but these are not 
relevant in the context of the study reported here, which is restricted to the minimal elastic anisotropy model described by the original de Gennes theory.

By contrast, active anchoring, has been reported only recently in the context of active matter, which refers to systems composed of interacting particles that generate motion through the supply of energy at the particle level~\cite{C7SM00325K, PhysRevLett.113.248303}. A range of biological systems including suspensions of swimming bacteria~\cite{Patteson2018, Wensink14308, yang2019quenching} belong to this class, which is characterized by novel intrinsically non-equilibrium phenomena with some degree of universality. Typically, active systems exhibit collective complex motion including turbulence at low Reynolds number. Interfaces of these systems, such as in active droplets and films, have been reported to exhibit instabilities, which are different from those of the bulk~\cite{C7SM00325K, PhysRevLett.113.248303, C9SM00859D}. The conditions under which these instabilities may lead to spontaneous motion of the bulk fluid are not yet completely understood. It is clear, however, that the properties of the interfaces of active nematics with passive fluids may be important in determining the fluid dynamics.
In particular, active anchoring has been reported as activity-induced flows produce a preferred alignment at the interface between an active nematic and a passive isotropic fluid. The interfacial alignment was found to be planar (director tangential to the interface) in active nematics where the particles produce extensile stresses (pushers), and homeotropic (director perpendicular to the interface) where the particles produce contractile stresses (pullers)~\cite{PhysRevLett.89.058101, MarchettiRMP2013, DellArciprete2018}. 

The equations for active nematic hydrodynamics have been proposed as a useful framework for studying active systems in 
general and bacterial colonies in particular~\cite{Doostmohammadi2018}. These are based on passive nematic hydrodynamics that uses the theory of de 
Gennes for the thermodynamics and Beris-Edwards equations for the dynamics. The activity (extensile or contractile) is 
added to the dynamics coupled linearly to the order parameter derivatives. In the models where active anchoring was reported, the nematic elasticity is controlled by a single elastic constant, either because the model is strictly two-dimensional or for simplicity 
when the three-dimensional de Gennes free energy is used. As a consequence the passive nematics considered in those
models have no interfacial anchoring. 

Here we introduced elastic anisotropy in the passive nematic model, at the level of the simplest de Gennes theory, and investigate the effects of the two types of interfacial anchoring in active nematics. We set the flow aligning parameter to zero to avoid a third bulk aligning mechanism and consider multiphase (MP) and multicomponent (MC) hydrodynamic models, the dynamics of which do not or do conserve the nematic order parameter, respectively. We start by analyzing the characteristic time for the relaxation of the directors at translational invariant interfaces for both types of anchoring. Then, we investigate the competition of these two effects for a wide range of parameters and find a threshold activity for which active anchoring prevails over the passive one, for both extensile and contractile active stresses. Next we consider two-dimensional interfaces and active nematic droplets and find that the dynamics is not affected by the elastic anisotropy except at extremely low activities at which the interfaces remain static. The elastic anisotropy, however, affects the active length, which becomes anisotropic.

This paper is organized as follows. In Sec.~\ref{method-sec} we describe the MP and MC models with elastic anisotropy, which are simulated using a hybrid method of lattice-Boltzmann and finite differences. In Sec.~\ref{passive-sec} we measure the relaxation time of the interfacial director in passive nematics with different elastic anisotropies. In Sec.~\ref{active-sec} we investigate the competition between the two types of interfacial anchoring. In Sec.~\ref{2d-sec} we simulate two-dimensional interfaces and analyze the effect of the elastic anisotropy on the dynamics of the system. Finally we summarize and conclude in Sec.~\ref{conc-sec}.

\section{Method}
\label{method-sec}

In this section, we describe briefly two models of nematic hydrodynamics used in this paper. The multiphase (MP) model that describes thermotropic liquid crystals does not conserve the nematic order parameter: the number of ordered particles is not fixed as the dynamics does not conserve the nematic order parameter. By contrast, the multicomponent (MC) model that describes lyotropic liquid crystals, conserves both the number of nematogen particles as well as the particles in the isotropic fluid. The nematic ordering is modelled by the $Q$ tensor de Gennes free energy with elastic anisotropy, i.e., with two elastic constants.   

\subsection{Multiphase model}
\label{mp-sec}

For uniaxial nematics, the nematic order may be described by the director field $n_\alpha$, which is the average direction of alignment, and the scalar order parameter $S$, which measures the degree of alignment. These two fields are combined in the tensor order parameter,  $Q_{\alpha \beta} = S(n_\alpha n_\beta - \delta_{\alpha \beta}/3)$, which is traceless and symmetric. The equilibrium state of the system is given by the minimum of the de Gennes free energy $\mathcal{F} = \int_V \,d^3 r\, f_{LdG}$, where the energy density is given by:
\begin{align}
 &f_{LdG}(\gamma) = \frac{A_0}{2}\left( 1- \frac{\gamma}{3} \right) Q_{\alpha \beta}^2 - \frac{A_0\gamma}{3} Q_{\alpha \beta} Q_{\beta \gamma} Q_{\gamma \alpha}  \nonumber \\
 &+ \frac{A_0\gamma}{4} (Q_{\alpha \beta} Q_{\alpha \beta} )^2  + \frac{L_1}{2} (\partial _\gamma Q_{\alpha \beta})^2 + \frac{L_2}{2}\partial_\epsilon Q_{\nu \epsilon} \partial_\gamma Q_{\nu\gamma}.
 \label{fldg-eq}
\end{align}
Here $A_0$ is a positive constant that sets the scale of the bulk free energy, $L_1$ and $L_2$ are the two elastic constants and $\gamma$ is the ordering field, e.g., a concentration related parameter in lyotropic liquid crystals~\cite{doi1988theory, beris1994thermodynamics}. At coexistence between the isotropic ($S=0$) and nematic ($S=S_N$) phases, which is set throughout the simulations, $\gamma$ is $2.7$ while the nematic order parameter is $S_N=1/3$. As reported by de Gennes, the sign of $L_2$ determines the interfacial anchoring at the nematic-isotropic interface and in what follows we call nematics with negative $L_2$ homeotropic and those with positive $L_2$ planar. Nematics with zero $L_2$ are elastically isotropic and do not exhibit interfacial anchoring. The uniaxial assumption of the $Q$ tensor is exact in the bulk and in homeotropic interfaces and a very good approximation in planar ones.

\begin{figure}[h]
\center
\includegraphics[width=\linewidth]{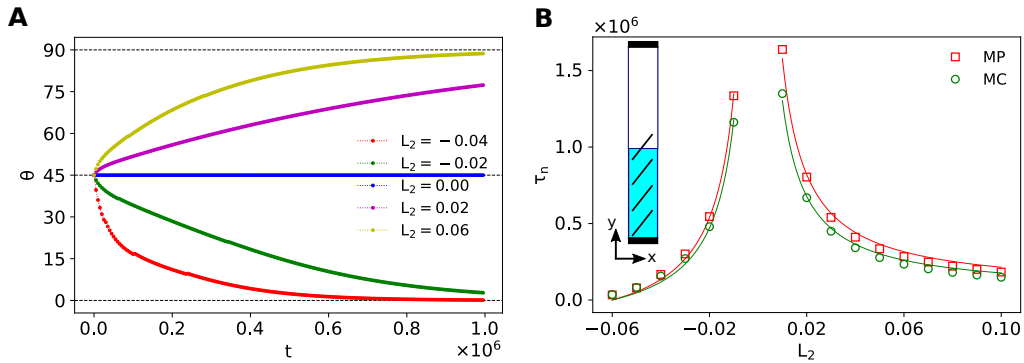}
\caption{ Relaxation of the director at a passive nematic interface. A) Angle $\theta$ between the director and the normal to the interface, $y$, for systems with different elastic anisotropies $\epsilon=L_2/L_1$ based on the multiphase model. For values of $L_2$ close to zero (e.g., $L_2=0.02$ as shown in the figure), the directors do not reach their equilibrium orientation in the simulation time, which diverges as $L_2$ vanishes. B) Relaxation time $\tau_n$ of the interfacial director as a function of $L_2$. The squares and circles are results from the simulations of the multiphase and multicomponent models respectively, while the solid lines are fits as described in the text. Nematics with positive $L_2$ are planar while those with negative $L_2$ are homeotropic. The inset illustrates schematically the 1D domain with parallel plates at the top and bottom. The domain is periodic allowing flow in the $x$ direction.}
\label{passive-fig}
\end{figure}

The time evolution of the nematic is governed by the Beris-Edwards equation~\cite{beris1994thermodynamics}, the continuity and the Navier-Stokes equation~\cite{beris1994thermodynamics, landau1987}, respectively:
\begin{align}
  &\partial _t Q_{\alpha \beta} + u _\gamma \partial _\gamma Q_{\alpha \beta} - S_{\alpha \beta}(W_{\alpha\beta}, D_{\alpha\beta}) = \Gamma H_{\alpha\beta} , \label{beris-edwards-eq} \\
  &\partial _\beta u_\beta = 0, \label{continuity-eq}\\
&\rho\partial_t  u_\alpha + \rho u_\beta \partial _\beta   u_\alpha  =  \partial_\beta [\eta(\partial_\alpha u_\beta + \partial_\beta u_\alpha)  + \sigma^{\text n}_{\alpha\beta} -\zeta Q_{\alpha\beta} ].  \label{navier-stokes-eq}
\end{align}
Eq.~\eqref{beris-edwards-eq} describes the evolution of the order parameter $Q_{\alpha\beta}$ while Eqs.~\eqref{continuity-eq} and \eqref{navier-stokes-eq} describe the dynamics of the velocity field $u_\alpha$. Here $\Gamma$ is the system dependent rotational diffusivity, $\rho$ is the density and $\eta$ is the shear-viscosity. The last term in Eq.~\eqref{navier-stokes-eq} is the active stress, which corresponds to a force dipole density, with $\zeta$ the activity parameter being positive for extensile stresses (systems composed by pushers) and negative for contractile ones (systems composed by pullers)~\cite{PhysRevLett.89.058101}. Gradients in $Q$ thus produce a flow field, which is the source of the hydrodynamic instabilities of active nematics. The co-rotational term is as follows: 
\begin{align}
 S_{\alpha \beta} =& ( \xi D_{\alpha \gamma} + W_{\alpha \gamma})\left(Q_{\beta\gamma} + \frac{\delta_{\beta\gamma}}{3} \right) 
 + \left( Q_{\alpha\gamma}+\frac{\delta_{\alpha\gamma}}{3} \right)(\xi D_{\gamma\beta}-W_{\gamma\beta}) \nonumber \\& 
 - 2\xi\left( Q_{\alpha\beta}+\frac{\delta_{\alpha\beta}}{3}  \right)(Q_{\gamma\epsilon} \partial _\gamma u_\epsilon), 
 \label{corrotational-eq}
\end{align}
where $W_{\alpha\beta}= (\partial _\beta u_\alpha - \partial _\alpha u_\beta )/2$ is the vorticity, $D_{\alpha\beta} = (\partial _\beta u_\alpha + \partial _\alpha u_\beta )/2$ is the shear rate
and $\xi$ is the flow alignment parameter, which characterizes the relative importance of the shear rate and the vorticity in the flow alignment of the particles.
The molecular field $H_{\alpha\beta}$ describes the relaxation of the order parameter towards equilibrium:  
\begin{align}
 H_{\alpha\beta} = -\frac{\delta \mathcal{F}}{\delta Q_{\alpha\beta}} + \frac{\delta_{\alpha\beta}}{3} \Tr \left( \frac{\delta \mathcal{F}}{\delta Q_{\gamma \epsilon}} \right).
\end{align}
The passive nematic stress tensor is~\cite{beris1994thermodynamics}:
\begin{align} 
 \sigma_{\alpha\beta}^{\text n} =& -P_0 \delta_{\alpha\beta} + 2\xi \left( Q_{\alpha\beta} +\frac{\delta_{\alpha\beta}}{3} \right)Q_{\gamma\epsilon}H_{\gamma\epsilon} - \xi H_{\alpha\gamma} \left( Q_{\gamma\beta}+\frac{\delta_{\gamma\beta}}{3} \right) - \xi \left( Q_{\alpha\gamma} +\frac{\delta_{\alpha\gamma}}{3} \right) H_{\gamma \beta} \nonumber \\ 
 & +\sigma^{\text{s}}_{\alpha\beta}  + Q_{\alpha\gamma}H_{\gamma\beta} - H_{\alpha\gamma}Q_{\gamma\beta} ,
 \label{passive-pressure-eq}
\end{align}
where $P_0$ is the isotropic pressure and 
\begin{align}
 \sigma^{\text{s}}_{\alpha\beta} =  - \frac{\delta \mathcal{F}}{\delta (\partial_\beta Q_{\gamma\nu})}\, \partial _\alpha Q_{\gamma\nu} .
\end{align}

This system of differential equations is solved using a hybrid method with the same spatial discretization: Eq.~\eqref{beris-edwards-eq} is solved using finite-differences and Eqs.~\eqref{continuity-eq} and \eqref{navier-stokes-eq} are recovered in the macroscopic limit with the lattice-Boltzmann method. The method is similar to those used in Refs.~\cite{C9SM00859D, refId0,C9SM02306B} and, for simplicity, we use the single relaxation time approximation in the Boltzmann equation~\cite{succi2018lattice, kruger2016lattice}. Our results are given in units such that the distance between nodes is $\Delta x = 1$ and the time step is $\Delta t = 1$ (lattice units). 

The parameters used in the simulations of the MP model, except when stated otherwise, are: $L_1=0.04$, $\xi=0$ (pure flow tumbling particles), $A_0=0.1$, $\rho=10$, $\tau=1.0$ (or, kinematic viscosity $\nu=(\tau-1/2)/3=0.17$) and $\Gamma=0.34$. The second elastic constant $L_2$ is varied in order to investigate the role of the elastic anisotropy $\epsilon=L_2/L_1$. Systems with positive $L_2$ are called planar and systems with negative $L_2$ homeotropic. Systems with $L_2=0$ are elastically isotropic. For values of $\epsilon<-1.5$ (or $L_2<-0.06$), the liquid crystal becomes unstable~\cite{doi:10.1080/02678298708086335}. 

\begin{figure}[h]
\center
\includegraphics[width=\linewidth]{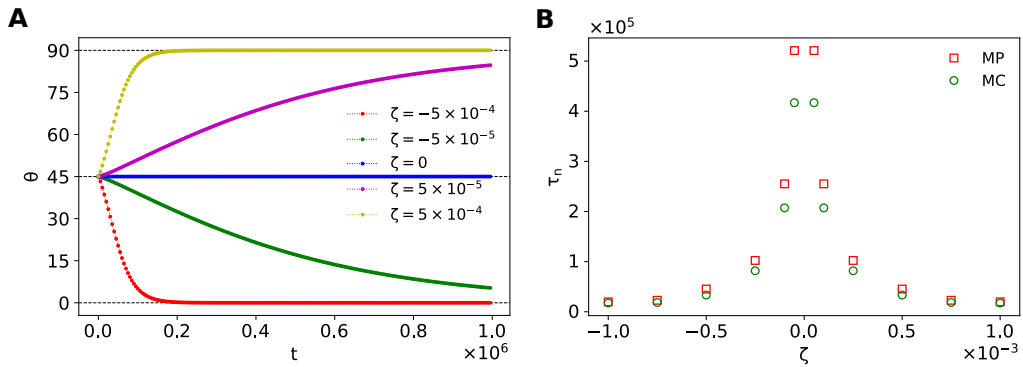}
\caption{Relaxation of the director at an active nematic interface, for an elastically isotropic system, $L_2=0$. A) Angle $\theta$ between the director and the normal to the interface, $y$, for systems with different activities $\zeta$ based on the multiphase model. Systems with positive $\zeta$ are extensile and those with negative $\zeta$ contractile. Extensile systems drive planar active anchoring while contractile ones drive homeotropic anchoring. We calculate the relaxation time as described in the text (e.g., $\tau_n=5.21 \times 10^5$ for $\zeta=-5\times 10^{-5}$ and $\zeta=5\times 10^{-5}$). B) Relaxation time $\tau_n$ of the interfacial director as a function of the activity $\zeta$. The squares and circles are the results from simulations of the multiphase and multicomponent models respectively.}
\label{angle-zeta-fig}
\end{figure}

\subsection{Multicomponent model}
\label{mc-sec}

The MC model considers two immiscible fluids the concentration of which is given by the scalar field $\phi$: $\phi=1$ for the nematic and $\phi=0$ for the isotropic fluid. It is usually employed for emulsions of nematogen particles in isotropic fluids. We consider the system in the two-phase region. The free energy density is written as follows:
\begin{align}
f &= \frac{a}{4}\phi^2(1-\phi)^2  + \frac{K}{2} \left(\partial_\gamma \phi \right)^2  + f_{LdG}(\gamma(\phi)),
\label{energy-mc-eq}
\end{align}
where $a$ is a positive constant, and $K$ is an elastic constant related to the surface tension, which penalizes inhomogeneities in the concentration field. $f_{LdG}(\gamma(\phi))$ is given by Eq.~\eqref{fldg-eq} where $\gamma$ is now a function of the concentration: $\gamma(\phi) = \gamma_0 + \gamma_s\phi$, with $\gamma_0$ and $\gamma_s$ being positive constants. The linear dependence accounts for the increase in nematic ordering with the concentration of particles in lyotropic liquid crystals. We set $\gamma_0=2.6$ and $\gamma_s=0.2$ which implies that the order is nematic when $\phi=1$ and isotropic when $\phi=0$ (recall that there is a first order nematic-isotropic transition at $\gamma = 2.7$).     

The dynamic equations are Eqs~\eqref{beris-edwards-eq}, \eqref{continuity-eq} and \eqref{navier-stokes-eq} and the Cahn-Hilliard equation:
\begin{eqnarray}
 &\partial_t \phi + \partial_\beta (\phi u_\beta) = M \nabla^2 \mu,
 \label{cahn-hilliard-eq}
\end{eqnarray}
where $M$ is the mobility constant that controls the diffusion of the concentration field and the chemical potential is:
\begin{eqnarray}
 \mu = \frac{\partial f}{\partial \phi}-\partial _\gamma \left[    \frac{\partial f}{\partial \left( \partial_\gamma \phi \right)}\right].
\end{eqnarray}
Eq.~\eqref{cahn-hilliard-eq} is solved with finite-differences. In Eq.~\eqref{passive-pressure-eq}, the term $\sigma^s_{\alpha\beta}$ becomes:
\begin{eqnarray}
 \sigma^s_{\alpha\beta} = \left( f-\mu \phi \right) \delta_{\alpha\beta} - \frac{\delta \mathcal{F}}{\delta \left(  \partial_\beta \phi \right)} \partial_\alpha \phi - \frac{\delta \mathcal{F}}{\delta \left(   \partial_\beta Q_{\gamma\nu}\right)}\partial_\alpha Q_{\gamma \nu}
\end{eqnarray}
This model is that used in Refs.~\cite{C6SM01275B, PhysRevE.74.041708} with the addition of a second elastic term in the free energy. 

We note that pure 2D nematics are elastically isotropic as the two elastic constants are degenerate in this theory. Indeed, in 2D, $(\partial_\gamma Q_{\alpha\beta})^2= 2 (\partial _\gamma S)^2$ and $\partial_\gamma Q_{\gamma \nu} \partial_\mu Q_{\mu\nu} = (\partial _\gamma S)^2$ while, in 3D, $(\partial_\gamma Q_{\alpha\beta})^2= 2/3 (\partial _\gamma S)^2$ and $\partial_\gamma Q_{\gamma \nu} \partial_\mu Q_{\mu\nu} = (\partial _\gamma S)^2/9 + (n_\gamma \partial_\gamma S)^2/3$. 

The parameters used in the simulations of the MC model, except when stated otherwise, are: $L_1=0.04$, $\xi=0$ (pure flow tumbling particles), $A_0=0.1$, $\rho=10$, $\tau=1.0$ (or, kinematic viscosity $\nu=(\tau-1/2)/3=0.17$), $\Gamma=0.34$, $K=0.08$, $a=0.05$ and $M=0.34$.

\section{Passive nematic}
\label{passive-sec}

In this section, we report the effect of elastic anisotropy in the director relaxation at the interface of passive nematics, using both the MP and the MC models. Stability requires $L_2>-3L_1/2$, which, for our choice of parameters becomes $L_2>-0.06$~\cite{doi:10.1080/15421407108082773}. 

Elastic anisotropy implies the existence of two bulk correlation lengths in the directions perpendicular and parallel to the nematic orientation. The normal and  parallel correlation lengths for the passive nematic are, respectively~\cite{doi:10.1080/15421407108082773}:
\begin{align}
 \ell_N^{\perp}= \sqrt{\frac{(3L_1+2L_2)}{ A_0(3-\gamma) }}, \quad \text{and} \quad \ell_N^{\parallel}= \sqrt{\frac{(3L_1+L_2/2)}{ A_0(3-\gamma) }}.
 \label{ln-norm-par-eq}
\end{align}
Note that, at the stability limit, $L_2=-0.06$, the normal correlation length vanishes $\ell_N^{\perp}=0$. The  simulation results are not reliable when $L_2$ is close to the stability limit as $\ell_N^\perp$ becomes comparable to the lattice spacing. 

We start by simulating effectively 1D systems with length $L_Y= 128$, periodic in the $x$ direction and with parallel solid plates (no-slip boundary condition) at the top and bottom as illustrated in the inset of Fig.~\ref{passive-fig}B. The nematic fills the bottom half of the box while the upper half is isotropic with an interface in the middle. The plate at the top is isotropic and that at the bottom is nematic with $S=S_N$ and zero director gradient (no preferred alignment). In addition, for the MC model, we set $\phi=0$ at the top and $\phi=1$ at the bottom. The directors are initialized at an angle $\theta=45^\circ$ with the $y$ axis (normal to the interface). Note that we are assuming translational invariance in the $x$ direction, which is a reasonable approximation for weakly active systems. At higher activities, the flat interface becomes unstable and the system exhibits active turbulence as discussed in Sec.~\ref{2d-sec}.     

As mentioned previously, $L_2<0$ favors homeotropic alignment at the interface ($\theta_{\text{eq}}=0$) while $L_2>0$ favors planar alignment ($\theta_{\text{eq}}=90^\circ$) as these configurations minimize the surface tension~\cite{doi:10.1080/15421407108082773}. In the absence of other aligning fields, as in the simulations reported in this section, this will be the interfacial anchoring in the equilibrium state. Fig.~\ref{passive-fig}A illustrates the time evolution of the director alignment at the interface (defined as the position where $S=S_N/2$). The equilibrium alignment depends on the sign of $L_2$ only, but the relaxation time depends strongly on the magnitude of $L_2$. The relaxation time is calculated through the fit:
\begin{align}
 \vert  \theta - \theta_{eq}  \vert = 45^\circ \exp\left( -\frac{t}{\tau_n}\right)
 \label{char-time-eq}
\end{align}
and is plotted in Fig.~\ref{passive-fig}B as a function of $L_2$. Dimensional analysis following Ref.~\cite{p1995physics},  suggests that the relaxation time scales as: $\tau_n\propto \ell_N^2/(\Gamma L_2)$. The nematic correlation length in the perpendicular direction to the interface depends on the sign of $L_2$ due to the interfacial anchoring ($\ell_N=\ell_N^\perp$ if $L_2<0$ and $\ell_N=\ell_N^\parallel$ if $L_2>0$). Using Eq.~\eqref{ln-norm-par-eq} we expect: $\tau_n =a_1\left(  \frac{L_1}{L_2}+\frac{2}{3} \right)$ for homeotropic nematics (when $L_2<0$) and $\tau_n = a_2\left( \frac{L_1}{L_2}+\frac{1}{6} \right)$ for planar ones (when $L_2>0$), where $a_1$ and $a_2$ are constants that may be obtained from a fit to $\tau_n$. These fits yield for the MP model, $a_1 = (-405 \pm 8)\times 10^3$ and $a_2 = (379 \pm 7)\times 10^3$ and, for the MC model, $a_1 = (-353 \pm 9)\times 10^3$ and $a_2 = (313 \pm 5)\times 10^3$. We note that $\tau_n$ tends to infinity as $L_2$ tends to zero, since as $L_2\rightarrow 0$ the elastic anisotropy vanishes and there is no preferred alignment at the interface.   

\begin{figure}[h]
\center
\includegraphics[width=0.8\linewidth]{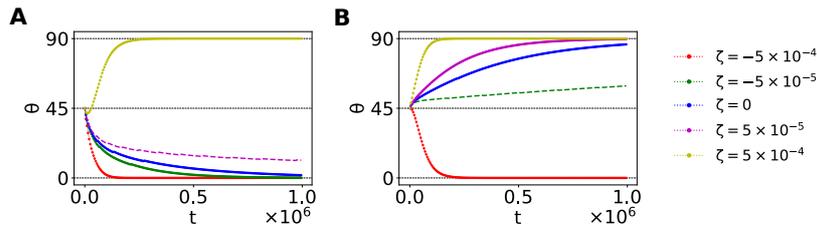}
\caption{Competition between the elastic anisotropy (controlled by $L_2$) and the activity in the multiphase model. The activity favors planar anchoring for extensile systems ($\zeta>0$) and homeotropic for contractile ones ($\zeta<0$). 
A) Passive homeotropic nematic with $L_2=-0.04$. B) Passive planar nematic with $L_2=0.04$. The different colors represent different activities (the same as in Fig.~\ref{angle-zeta-fig}A, where $L_2=0$). The curves for which there is competition and 
the effect of the anisotropy is stronger than that of active anchoring are represented with a dashed line with the color given in the legend. Notice that this only occurs at small activities ($\vert \zeta \vert \sim 10^{-5}$).}

\label{angle-competition-fig}
\end{figure}


\section{Active nematic}
\label{active-sec}

In Sec.~\ref{passive-sec}, we reported the effect of the elastic anisotropy on the interfacial anchoring. Here, we analyze the effect of activity, known as active anchoring, and the competition between these interfacial aligning mechanisms. We assume the particles are pure flow tumbling ($\xi=0$) and focus on the competition between elastic anisotropy and active anchoring. The effect of flow alignment is considered briefly at the end of this section.

Active anchoring has been reported previously~\cite{PhysRevLett.113.248303, C7SM00325K, C9SM00859D} for nematics with no elastic anisotropy. 
We consider a setup that is effectively 1D (invariant in the $x$ direction) with open boundary for the directors at the bottom. We start our analysis by assuming that the gradient of $S$ at the interface is much larger than the gradient of the director field.
As in previous sections we take the normal to the interface $\mathbf{m}$ in the $y$ direction and the direction parallel to the interface $\mathbf{l}$ is in the $x$ direction. The directors $\mathbf{n}$ are initialized at an angle of $45^\circ$ with the y-axis. 
Under these conditions, the active force $\mathbf{F}^{\text{a}} = -\zeta \nabla \cdot \mathbf{Q}$ may be written in terms of  parallel and normal components to the interface, given by:
\begin{align}
 \mathbf{F}_{\parallel}^{\text{a}} &= \zeta \vert \nabla S \vert (\mathbf{m}\cdot \mathbf{n})(\mathbf{l}\cdot \mathbf{n}) \mathbf{l} \label{act-force-par-eq}\\
  \mathbf{F}_\bot^{\text{a}} &= \zeta \vert \nabla S\vert \left[  (\mathbf{m}\cdot\mathbf{n})^2-\frac{1}{3}  \right]\mathbf{m}.\label{act-force-perp-eq}
\end{align}
This force acts close to the interface only due to its dependence on $\nabla S$. The perpendicular component of the force $\mathbf{F}_\bot^{\text{a}}$ acts towards the nematic for extensile systems, $\zeta>0$, and destabilizes a flat interface if it is strong compared to the surface tension~\cite{C7SM00325K, C9SM00859D}. Two dimensional systems at higher activities will be considered in Sec.~\ref{2d-sec} but here we restrict the analysis to systems at low activities, where translational invariance may be assumed.

\begin{figure}[h]
\center
\includegraphics[width=0.5\linewidth]{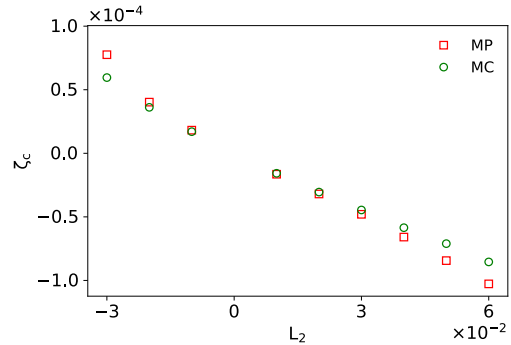}
\caption{Threshold activity $\zeta_c$ beyond which the active anchoring dominates the elastic anisotropy, as a function of $L_2$ in the multiphase and multicomponent models. This relation becomes non-linear close to the stability limit $L_2=-0.06$.  }
\label{l2-zeta-fig}
\end{figure}

The parallel active force is zero both for homeotropic and planar interfacial alignment. Both configurations are possible (see Eq.~\eqref{act-force-par-eq}) but one of them, which depends on the sign of $\zeta$, is unstable: for extensile systems, $\zeta>0$, the planar configuration is stable while the homeotropic one is unstable (for contractile systems, $\zeta<0$, the opposite applies). Notice that there is no flow in the $y$ direction due to the confinement.
In Fig.~\ref{angle-zeta-fig}A we plot the time evolution of the angle between the director and the normal to the interface ($y$ axis) 
for different activities and zero elastic anisotropy $L_2=0$. Homeotropic alignment is favored when $\zeta<0$ (contractile systems) and planar alignment is favored when $\zeta>0$, as expected.  The characteristic relaxation time, calculated as in Sec.~\ref{passive-sec}, depends on the activity as shown in Fig.~\ref{angle-zeta-fig}B. We note that $\tau_n$ for activities that destabilize the flat interface ($\zeta>0.001$, as discussed in Sec.~\ref{2d-sec}), is much smaller than the relaxation time associated with the elastic anisotropy. This suggests that active anchoring will override the anchoring driven by the elastic anisotropy in this range of activities. Note that $\tau_n$ diverges at the activity where there is no preferred anchoring, $\zeta=0$ with $L_2=0$. 

Figure~\ref{angle-competition-fig} illustrates the competition of these aligning effects for planar and homeotropic nematics (positive and negative $L_2$) and different activities. The two types of anchoring compete when the signs of $L_2$ and $\zeta$ are different.
At low activities $\zeta$, the alignment is driven by the elastic anisotropy while at higher activities $\zeta$, active anchoring is the effective aligning mechanism both for extensile and contractile systems. The dashed lines correspond to systems with antagonistic anchorings for which the elastic anisotropy prevails. We notice that, at higher activities ($\vert \zeta \vert=5\times 10^{-4}$), the characteristic times depend weakly on the elastic anisotropy. When the anchorings are agonistic, the characteristic times are greatly reduced.

We proceed to determine the activity where the interfacial anchoring mechanisms balance, $\zeta_c$, in antagonistic systems. 
At each value of the elastic anisotropy $L_2$, we search iteratively the value of $\zeta$ that yields a steady state anchoring close 
to the initial orientation. When the two effects balance, the characteristic time tends to infinity, so we analyze the directors at $5\times 10^{6}$ iterations, which is one order of magnitude longer than the larger $\tau_n$ reported in Fig.~\ref{angle-zeta-fig}. The value of $\zeta_c$ is found through cubic interpolation as the value of $\zeta$ that corresponds to $\theta=45^\circ$. In other words, if $\zeta=\zeta_c$, the anchoring remains at the initial value as the two aligning mechanisms are balanced. 
The result is plotted in Fig.~\ref{l2-zeta-fig} for both models, which reveals an approximately linear relation. Note that both extensile and contractile systems exhibit a similar behaviour. We omitted values of $L_2$ for which the nematic correlation length is smaller than the lattice spacing.

\begin{figure}[h]
\center
\includegraphics[width=\linewidth]{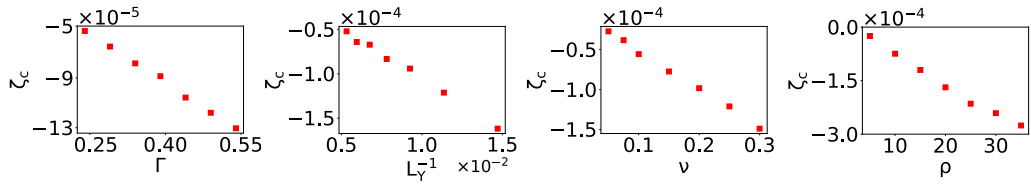}
\caption{Dependence of the threshold activity $\zeta_c$ on the other parameters in Eq.~\eqref{zetac-eq} for the multiphase model and fixed $L_2=0.05$ (planar interfacial anchoring). $\zeta_c$ is negative in this case (homeotropic active anchoring) to balance the effect of the elastic anisotropy. }
\label{parameters-fig}
\end{figure}
\begin{figure}[h]
\center
\includegraphics[width=0.5\linewidth]{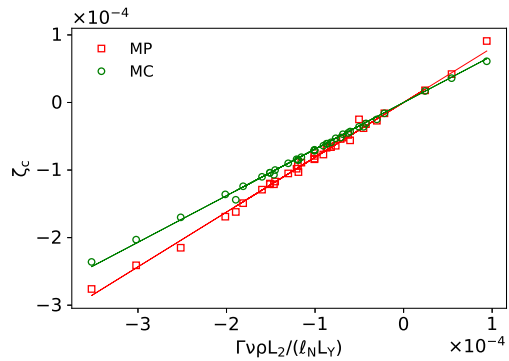}
\caption{Threshold activity $\zeta_c$ where the active anchoring prevails over the effect of elastic anisotropy as a function of a combination of the model parameters. The squares and circles are the results from simulations of the multiphase and the multicomponent models while the solid lines are linear fits. }
\label{zeta-all-fig}
\end{figure}

The threshold value of $\zeta_c$ plotted in Fig.~\ref{l2-zeta-fig} depends on the values of the other parameters of the model. To see this
we analyze the equations of motion (Eqs~\eqref{beris-edwards-eq} and \eqref{navier-stokes-eq}). The parallel component of the active force, Eq.~\eqref{act-force-par-eq}, is proportional to $\zeta$ and acts only close to the interface. As the system is confined between top and bottom plates, this force drives the flow in the $x$-direction with maximum velocity in the center (see the inset of Fig.~\ref{passive-fig}B). From Eqs.~\eqref{navier-stokes-eq} and \eqref{act-force-par-eq} we find that the $x$-component of the velocity field is $u_x \propto \zeta \vert \nabla S\vert /(\rho \nu)$ and the non-zero component of the vorticity field is $W_{xy}=-W_{yx} \propto \zeta \vert \nabla S \vert L_Y/(\rho \nu)$ at a given position $y/L_Y$. If there is no flow alignment, $\xi=0$, as we consider in most simulations, the co-rotational term in Eq.~\eqref{corrotational-eq} is proportional to the vorticity field with the non-zero component being $S_{xy} = - S_{yx} = W_{xy}(Q_{yy}-Q_{xx}) \propto \zeta \vert \nabla S \vert L_Y/(\rho \nu) $. In addition, the $xy$ component of the molecular field is $H_{xy} \propto (L_1 + L_2/2) \partial _y^2 Q_{xy}$. Thus, Eq.~\eqref{beris-edwards-eq} for the steady state gives:
\begin{align}
 \zeta_c \propto -\frac{\Gamma \rho \nu L_2}{\ell_N L_Y},
 \label{zetac-eq}
\end{align}
where we considered $\vert \nabla S \vert \approx S/\ell_N$ with the nematic correlation length $\ell_N$ being defined in Sec.~\ref{passive-sec}. This equation describes the linear relation between $\zeta_c$ and $L_2$ observed in Fig.~\ref{l2-zeta-fig}. Note that this relation holds both for extensile and contractile systems, e.g. if $L_2>0$, then $\zeta_c<0$ and vice-versa in line with the results of Fig.~\ref{l2-zeta-fig}.

\begin{figure}[h]
\center
\includegraphics[width=0.5\linewidth]{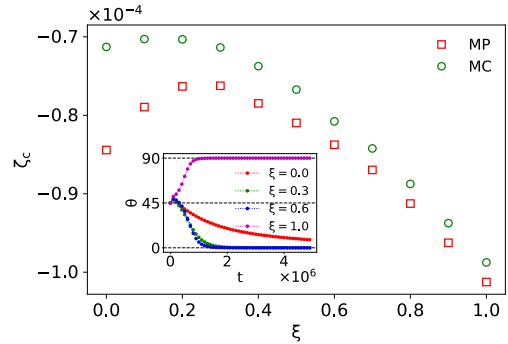}
\caption{Threshold activity $\zeta_c$ as a function of the alignment parameter $\xi$ for multiphase and multicomponent models with $L_2=0.05$.The inset shows the evolution of the initial anchoring at $45^\circ$ for the multiphase model and different values of $\xi$ for a contractile system with $\zeta=-9\times 10^{-5}$.} 
\label{zeta-xi-fig}
\end{figure}

In order to check the relation between $\zeta_c$ and the other parameters in Eq.~\eqref{zetac-eq} we performed simulations by changing one parameter at a time with the others kept fixed. We set $L_2=0.05$, which induces planar anchoring and thus $\zeta_c$ is negative (contractile). Fig.~\ref{parameters-fig} confirms that $\zeta_c$ depends linearly on each of these parameter as predicted by Eq.~\eqref{zetac-eq}. Furthermore, the results collapse on a straight line as shown in Fig.~\ref{zeta-all-fig}. Note that although the relation is linear for both models the slope differs: for the MP model it is $0.811 \pm 0.007$ while for the MC model it is $0.690 \pm 0.004$, where the errors come from the linear fit.

A generalization of Eq.~\eqref{zetac-eq} gives a dimensionless quantity that characterizes the relative efect of the elastic anisotropy and of the active anchoring:
\begin{align}
 An = \frac{\Gamma \rho \nu L_2}{\zeta \ell_N \ell_F},
 \label{number-ae-eq}
\end{align}
where $\ell_F$ is the characteristic length of the flow, i.e., the length scale over which the velocity field changes significantly. If $An>An^\ast$, the effect of the elastic anisotropy dominates over that of the active anchoring and the opposite holds if $An<An^\ast$, where $An^\ast$ is the value at threshold, when these effects balance. Our previous analysis yields for the thresholds $An^\ast=1.2$ in the MP model and $An^\ast=1.4$ in the MC model.

Finally, in Fig.~\ref{zeta-xi-fig} we briefly report the effect of the flow alignment parameter $\xi$ on $\zeta_c$ for a planar system with elastic anisotropy $L_2=0.05$ and all the other parameters as in Sec.~\ref{method-sec}. 
Flow alignment is a third mechanism which affects the bulk and consequently the interfacial alignment. Unlike the other mechanisms that act close to the interface flow alignment acts throughout the whole system. The existence of a peak is an interesting feature, which is, however, not fully understood. It is probably related to the fact that the particles are flow tumbling if $\xi<\xi^\ast$ and flow aligning if $\xi>\xi^\ast$, where $\xi^\ast=3S/(S+2)$. Thus, for the MP model $\xi^\ast = 0.43$ and, while for the MC model, $\xi^\ast =0.51$ since $\gamma=2.8$ and $S_N=0.41$. However, even in passive nematics flow alignment may lead to non-trivial effects~\cite{batista2014effect} and how it affects the interfacial anchoring of active nematics is beyond the scope of this work.

\section{Two-dimensions}
\label{2d-sec}

\subsection{Flat interface}

We proceed to analyze the effect of perturbations on the interfacial anchoring at higher activities, in particular the stability of the flat interface. As reported in Sec.~\ref{active-sec}, the effect of the elastic anisotropy on the interfacial anchoring is only observed at very low activities. The activities $\zeta$ considered in this and the following sections, are always positive (extensile systems) and above $\zeta_c$, i.e. active anchoring prevails and the interfacial anchoring is planar (even though the passive nematic is homeotropic).
\begin{figure}[h]
\center
\includegraphics[width=0.7\linewidth]{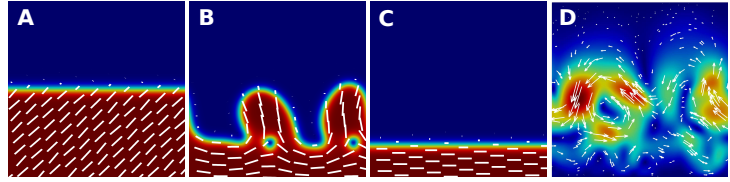}
\caption{Evolution of perturbations on initially flat interfaces in the multiphase model for a system with antagonistic anchoring mechanisms: homeotropic elastic anisotropy, $L_2=-0.02$, and planar active anchoring, $\zeta=0.005$ (above the threshold 
activity $\zeta_c$).
The figures depict different times steps: A) $t=0$, B) $t=5\times 10^4$, C) $t=5\times 10^{6}$. The colors represent the scalar order parameter with red being nematic and blue isotropic while the lines represent the directors. D) Velocity field corresponding to $t=5\times 10^4$, where the maximum velocity (red) is $\vert \mathbf{u} \vert=0.0033$. }
\label{2dmp-unst-fig}
\end{figure}

We consider square 2D simulation boxes, of size $L_X \times L_Y = 128 \times 128$, and systems with the same parameters as described in Sec.~\ref{method-sec}. Initially, the directors are set at $\theta=45^\circ$ with the $y$ axis, with random perturbations $\pm 5^\circ$ throughout (as shown in Fig.~\ref{2dmp-unst-fig}A). Although the directors are not constrained they remain in the $xy$-plane during the simulations. The perturbations generate active forces due to the gradients of $Q_{\alpha\beta}$ at the interface. If the active forces are sufficiently large, they will produce spontaneous flows that destabilize the flat interface. Thus the assumption of translational invariance made in the previous sections is no longer valid. The spontaneous flows have different effects in the two models considered in this work, as described below.

\begin{figure}[h]
\center
\includegraphics[width=0.7\linewidth]{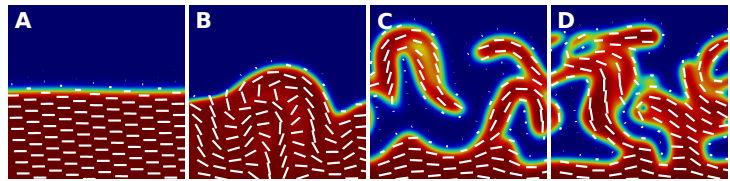}
\caption{Evolution of perturbations on initially flat interfaces in the multicomponent model at $t=5\times 10^6$ for extensile systems with $\zeta >  \zeta_c$ (active anchoring prevails). 
A) Agonistic extensile system (both mechanisms promote interfacial planar alignment) with $L_2=0.05$ and $\zeta=0.002$ and  C) $L_2=0.02$ and $\zeta=0.007$. Antagonistic extensile systems (homeotropic elastic anisotropy and planar active anchoring)  B) $L_2=-0.05$ and $\zeta=0.002$, and D) $L_2=-0.02$ and $\zeta=0.007$. The colors represent the scalar order parameter with red being nematic and blue isotropic while the lines represent the directors. }
\label{2dmc-unst-fig}
\end{figure}

The spontaneous flow in active nematics results from an interplay of elastic and active forces the ratio of which may be used to define an active length. In the one elastic constant approximation, the active length is given by $\ell_A=\sqrt{L_1/\zeta}$ (see e.g. Ref.~\cite{C9SM02306B}). In systems with elastic anisotropy, the active length becomes anisotropic as the elastic force is different in directions perpendicular and parallel to the local director, and thus we define two active lengths:  
\begin{align}
 \ell_A^\perp=\sqrt{\frac{L_1+2L_2/3}{\zeta}}\quad \text{and}\quad \ell_A^{\parallel}=\sqrt{\frac{L_1+L_2/6}{\zeta}}.
\end{align}
When the local director changes rapidly in space, as in chaotic or turbulent flows, we consider the geometrical average active length $\bar{\ell}_A = \sqrt{\ell_A^\perp \ell_A^{\parallel} }$.

In the MP model, at low activities the steady-state interface is flat as in \ref{2dmp-unst-fig}C. Even when the anchorings are antagonistic the perturbations in the interfacial director of extensile systems grow but eventually die out, through a complex dynamical transient. In the steady state, however, the interface is flat with uniform alignment driven by the active anchoring. The MP model dynamics does not conserve the nematic order parameter, and the amount of nematic in the steady state is usually different (smaller) than that in the initial state. The reorganization of the local nematic order driven by the active dynamics favors the isotropic phase, and this increases as the activity increases. In extensile systems the normal component of the active force, Eq.~\ref{act-force-perp-eq}, acts towards the nematic phase.
Note that this result differs from the results reported in Ref.~\cite{C9SM02306B}, where a similar model was used to describe the motion of interfaces of bacterial swarms.The main difference being that in the swarm model the particles are deep in the flow alignment regime ($\xi=0.7$) while here they are flow tumbling, $\xi=0$. Flow alignment favors the nematic phase~\cite{ThampiEPL2015} and it appears to affect the interfacial anchoring as reported in Sec.~\ref{active-sec}. 

In the MC model, the dynamics conserves both the isotropic and the nematic phases leading to marked differences in the interfacial behaviour. For agonistic extensile systems at low activities (both mechanisms promote interfacial planar alignment) the interface is flat, as shown in Fig.~\ref{2dmc-unst-fig}A for  $L_2=0.05$ and $\zeta=0.002$. This is not the case, however, at larger activities as illustrated in Fig.~\ref{2dmc-unst-fig}C for $L_2=0.02$ and $\zeta=0.007$. 
By contrast, in antagonistic extensile systems (homeotropic elastic anisotropy and planar active anchoring) the steady state interface is never flat, exhibiting undulations at low activities as illustrated in Fig.~\ref{2dmc-unst-fig}B) for $L_2=-0.05$ and $\zeta=0.002$, and extremely rough interfaces as shown in Fig.~\ref{2dmc-unst-fig}D for $L_2=-0.02$ and $\zeta=0.007$. 
In all cases, the active length is much smaller than the lateral dimensions of the simulation box, ranging from  $\bar{\ell}_A=5.5$ in Fig.~\ref{2dmc-unst-fig}A to $\bar{\ell}_A =2.1$ in Fig.~\ref{2dmc-unst-fig}D, where the interface has almost disappeared as anisotropic droplets of active nematic are dispersed in the isotropic fluid.
 
\subsection{Active nematic droplet}

Finally, we investigate the interfacial anchoring dynamics of an active droplet, using the MC model. 
We start with a nematic circular droplet of radius $R=50$ with directors aligned vertically. The box is of size $L_X \times L_Y = 200 \times 400$ and periodic boundary conditions are applied in both directions.
The other model parameters are those described in Sec.~\ref{method-sec}.  

\begin{figure}[h]
\center
\includegraphics[width=0.7\linewidth]{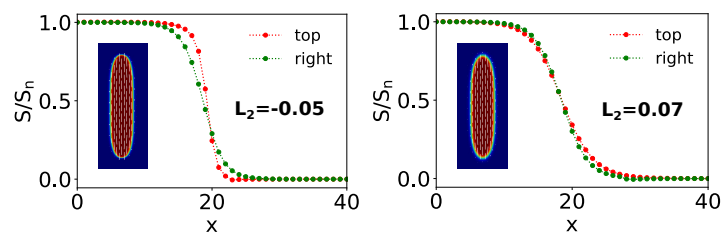}
\caption{Order parameter at the top and lateral interfaces of an active nematic droplet at low activity ($\zeta = 0.001$) using the MC model. The directors are mostly vertical and thus the alignment at the top and lateral interfaces is homeotropic and planar, respectively.}
\label{charuto-fig}
\end{figure}

The active force in the perpendicular direction, Eq.~\ref{act-force-perp-eq},  is extensile pointing outwards at the top and bottom as the directors are parallel to the interface normal and pointing inwards at the sides as the directors are perpendicular to the interface normal. 
Thus, the droplet is expected to elongate in the vertical direction. In Fig.~\ref{charuto-fig}, the insets depict the droplet at low activity ($\zeta=0.001$) in the steady state at two values of the elastic anisotropy, $L_2$ ($-0.05$ and $0.07$). At low activities, the surface tension balances the active force and the droplet is static and elongated in the vertical direction as expected~\cite{PhysRevLett.112.147802}. There are minor differences in the steady state droplets when we change the elastic anisotropy, $L_2$. We notice that, due to the anisotropic correlation length, the interfacial width varies along the droplet interface. At the top and bottom, where the directors are homeotropic to the interface, the interfacial width is different from that at the sides, where the directors are planar. This effect is visible in Fig.~\ref{charuto-fig}, where the scalar order parameter profile at the top and at the side of the droplet is plotted. This difference is largest for antagonistic systems, in line with the expressions for the correlation length.

\begin{figure}[h]
\center
\includegraphics[width=0.6\linewidth]{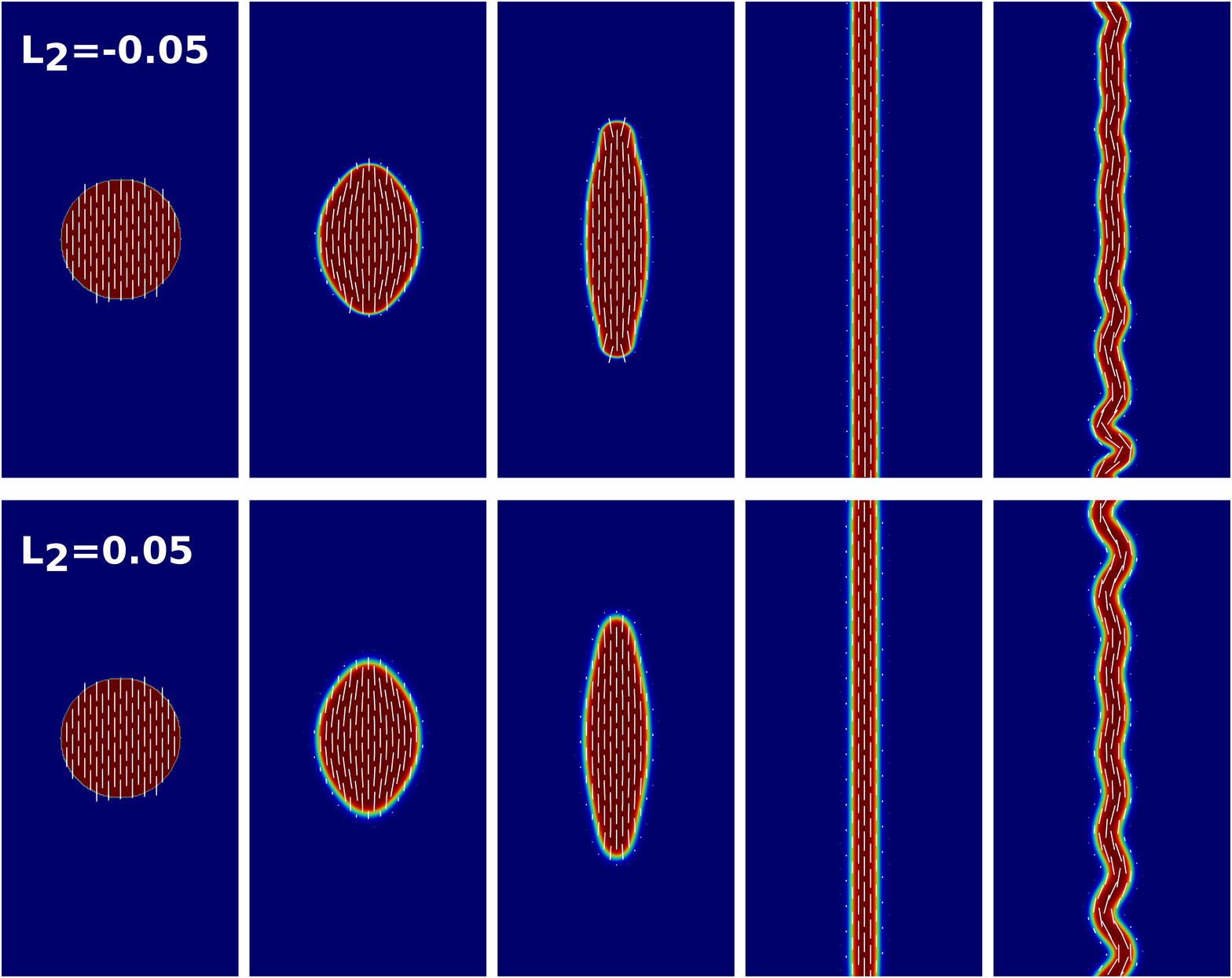}
\caption{Time evolution of an active nematic droplet at $\zeta = 0.01$ using the multicomponent model at different elastic anisotropies, $L_2$. }
\label{expanding-fig}
\end{figure}

\begin{figure}[h]
\center
\includegraphics[width=0.8\linewidth]{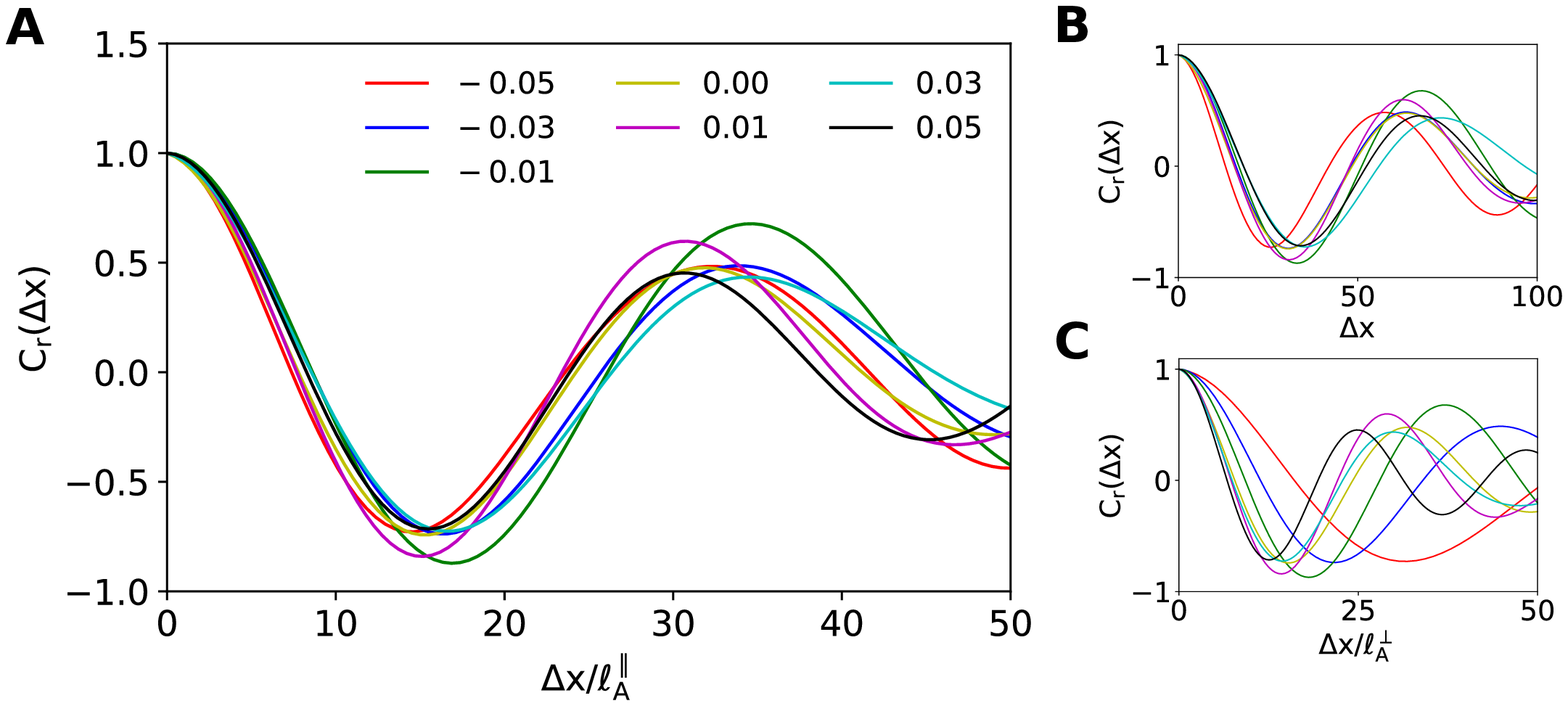}
\caption{Spatial correlation function of the interfacial height of the active nematic stripe. A) Distances scaled by the parallel active length $\ell_A^\parallel$. B) Original functions (not scaled). C) Distances scaled by the normal active length $\ell_A^{\perp}$.}
\label{correlation-fig}
\end{figure}

At higher activities, the droplet stretches until it reaches the limits of the periodic domain where it becomes a nematic stripe as shown in Fig.~\ref{expanding-fig} for two values of the elastic anisotropy, $L_2$. There is no significant change in the dynamics when the elastic anisotropy $L_2$ is varied. The dynamics of the stretching droplet and of the interfacial alignment are dictated by 
the activity.  The nematic stripe in Fig.~\ref{expanding-fig} becomes unstable and undulates due to the bend instability of extensile active nematics~\cite{PhysRevLett.113.248303}. The period of the undulations depends on the active length, as they are caused by active forces and balanced by elastic ones. We measured the spatial correlation function of the interface height at different elastic anisotropies $L_2$ (fixed $L_1$) as the undulation instability sets in, defined as the time when the standard deviation of the interface position is $\sigma_y=1$. This occurs at different times depending on $L_2$ (at $t=69000$ for $L_2=-0.05$ and $t=97000$ for $L_2=0.06$). As the active forces prevail, the parallel active length is used to collapse the curves in Fig.~\ref{correlation-fig}A. Figs.~\ref{correlation-fig}B and C show the original correlation functions and a collapse attempt using the perpendicular active length confirming that these are less adequate choices. 

In summary, the effect of the elastic anisotropy in active nematics is revealed through the anisotropy of the active length. It plays no significant role in the dynamics or in the interfacial alignment as the active anchoring is much stronger. Thus it is safe to combine them in a one elastic constant approximation.

\section{Conclusion}
\label{conc-sec}

We have investigated the effect of elastic anisotropy in the interfacial alignment of active nematics, at the passive nematic-isotropic coexistence. We introduced a second elastic constant in the free energy of a multiphase and a multicomponent model. We calculated the relaxation time of the interfacial directors as a function of the elastic anisotropy. A similar calculation was carried out for an elastically isotropic active nematic as a function of the activity. Using simulations and analytical arguments, we found the threshold activity at which active anchoring prevails over the effect of elastic anisotropy in setting the interfacial anchoring. This was found to depend linearly on the second elastic constant and the other model parameters, for both contractile and extensile systems. In addition, we investigated the effect of fluctuations on the stability of a flat interface of an extensile active nematic and found that in the multiphase model the steady state interface is flat while in the multicomponent model the flat interface is unstable and breaks up at high activities. Simulations of a stretching active nematic droplet revealed that the elastic anisotropy does not affect significantly its time evolution. It affects, however, the active length, which depends on the director field orientation and becomes anisotropic. We conclude that the interfacial aligning effects of activity are in general much stronger than the effects of the elastic anisotropy.

\vskip6pt

\enlargethispage{20pt}



\competing{The author(s) declare that they have no competing interests.}

\funding{We acknowledge financial support from the Portuguese Foundation for Science and Technology (FCT) under the contracts: PTDC/FIS-MAC/28146/2017 (LISBOA-01-0145-FEDER-028146), UIDB/00618/2020 and UIDP/00618/2020.}




%
%
%
%
%

\bibliography{rsc} 
\bibliographystyle{RS} 

\end{document}